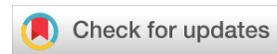

# REVISED On the evaluation of research software: the CDUR procedure [version 2; peer review: 2 approved]

Teresa Gomez-Diaz [ID]1, Tomas Recio [ID]2

1Laboratoire d'Informatique Gaspard-Monge, Centre National de la Recherche Scientifique, University of Paris-Est Marne-la-Vallée, Marne-la-Vallée, France
2Universidad de Cantabria, Santander, Spain

## Abstract

**Background:** Evaluation of the quality of research software is a challenging and relevant issue, still not sufficiently addressed by the scientific community.

**Methods:** Our contribution begins by defining, precisely but widely enough, the notions of research software and of its authors followed by a study of the evaluation issues, as the basis for the proposition of a sound assessment protocol: the CDUR procedure.

**Results:** CDUR comprises four steps introduced as follows: **C**itation, to deal with correct RS identification, **D**issemination, to measure good dissemination practices, **U**se, devoted to the evaluation of usability aspects, and **R**esearch, to assess the impact of the scientific work.

**Conclusions:** Some conclusions and recommendations are finally included. The evaluation of research is the keystone to boost the evolution of the Open Science policies and practices. It is as well our belief that research software evaluation is a fundamental step to induce better research software practices and, thus, a step towards more efficient science.

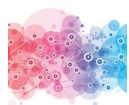

This article is included in the Research on Research, Policy & Culture gateway.

## Open Peer Review

**Reviewer Status** ✓ ✓

|  | Invited Reviewers | |
|---|---|---|
|  | **1** | **2** |
| **version 2** (revision) 26 Nov 2019 | ✓ report | |
|  | ↑ | |
| **version 1** 05 Aug 2019 | ? report | ✓ report |

1. **Francisco Queiroz** [ID], University of Leeds, Leeds, UK

2. **Jean-Pierre Merlet** [ID], Inria, Sophia-Antipolis, Valbonne, France

Any reports and responses or comments on the article can be found at the end of the article.

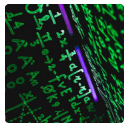 This article is included in the Mathematical, Physical, and Computational Sciences collection.

**Corresponding authors:** Teresa Gomez-Diaz (teresa.gomez-diaz@univ-mlv.fr), Tomas Recio (tomas.recio@unican.es)

**Author roles: Gomez-Diaz T**: Conceptualization, Formal Analysis, Funding Acquisition, Investigation, Methodology, Project Administration, Supervision, Validation, Visualization, Writing – Original Draft Preparation, Writing – Review & Editing; **Recio T**: Conceptualization, Formal Analysis, Investigation, Methodology, Project Administration, Supervision, Validation, Visualization, Writing – Original Draft Preparation, Writing – Review & Editing

**Competing interests:** No competing interests were disclosed.

**Grant information:** Publication of this article is supported by the Gaspard-Monge computer science laboratory (LIGM) at the University of Paris-Est Marne-la-Vallée.
*The funders had no role in study design, data collection and analysis, decision to publish, or preparation of the manuscript.*

**How to cite this article:** Gomez-Diaz T and Recio T. **On the evaluation of research software: the CDUR procedure [version 2; peer review: 2 approved]** F1000Research 2019, **8**:1353 https://doi.org/10.12688/f1000research.19994.2

**First published:** 05 Aug 2019, **8**:1353 https://doi.org/10.12688/f1000research.19994.1

## 1 Introduction

Scientific software is a key component in today's science and engineering development [1]. As described in detail in the Research software definition section a particular, yet fundamental, subset of scientific software is the research software (RS)[i] that is developed and used by researchers in the process of doing scientific research in public institutions or publicly funded projects[2].

Computer software was included among the positive spillovers of knowledge production in [2] (p.508):

> *"What is alluded to here is that there may be important positive spillovers across projects in the form of 'learning effects'. [...] which often remain in the region of tacit knowledge [...] including the development of generic computer software for performing data processing, storage, retrieval and network transmission".*

Henceforth, its importance to the scientific enterprise is generally accepted [3]:

> *"Modern science depends on software. Software analyzes data, simulates the physical world, and visualizes the results; just about every step of scientific work is affected by software"*

Similarly, most of the cited references in this work highlight the central role of software development in science nowadays. As this central role is increasingly assumed, it is also noticeable the emergence of some serious drawbacks. For example, finding scientific software can be a quite hard enterprise [4,5]; and difficulties can also arise when dealing with software citations [6,7]. Moreover, in [3] we can find:

> *"Software is increasingly important to the scientific enterprise, and science-funding agencies are increasingly funding software work. Accordingly, many different participants need insight into how to understand the relationship between software, its development, its use, and its scientific impact."*

Research quality evaluation is an intrinsically embedded component of research itself, with deep impact ranging from the enhancement of academic careers to boosting new knowledge production [8]. Accordingly, in the Key evaluation issues section we discuss and develop the intricate notion of evaluation in the context of research software, considering both the perspective of the evaluators and of the evaluated researchers. In particular, we clarify in the CDUR proposal section whether we are evaluating research, software, or research software as a scientific output. Likewise, in the same section we settle the basis to decide, within a research software evaluation scheme, when and how we are evaluating some software or its associated research.

Our goal is then to set up a basis for a simplified and flexible protocol concerning research software evaluation: what we have named as the CDUR procedure. It includes four stages, labelled as **C**itation, **D**issemination, **U**se and **R**esearch, that are thoroughly developed in the CDUR proposal section. The procedure is meant to help all the key actors involved in the evaluation process, and it applies to any scientific area, as the considered research software aspects are quite transversal. CDUR will provide insight in the relationship between software, its development, its use, and its scientific impact, as we will show in this work.

We are aware that there are plenty of references in specialized journals regarding software quality, free/open source or educational software assessment methodologies (e.g. the related Wikipedia pages or some of the Software Sustainability Institute (SSI) documents[3]). The testing, validation and verification of such types of software are well established concepts that proceed in a direction that we are not going to pursue in this work. There are also different publications concerning research evaluation (e.g. [8] for a recent study with more than 50 references on the subject) and a very complete review on the literature for scientific software testing can be found in [1].

Moreover, in the context of Open Science career assessment, the European Commission report [9] considers as an evaluation criteria the use and the development of free/open source software, although without proposing any concrete method for achieving such criterium; and similar considerations can be found in the NASA report concerning open source software policies and career credit [10] (p.74). Besides, the importance of the evaluation step in the context of future scholarly communication and Open Science is stressed in [11]: *"The conclusion is actually simple: the evaluation of research is the keystone"*.

Nevertheless, we have not been able to find publications addressing, in particular, evaluation of software developed for scientific research, (not scientific software in general), or concerning the evaluation of research software as a whole process (and not just testing). Thus, in our opinion, there is a clear need to approach the issue we are dealing with in this paper, concerning a more precisely determined object (research software) although in a wider evaluation context (as a global process).

Our contributions are distributed along this article as follows: next, the Research software section is devoted to discuss the different aspects related to the concept of research software and its associated issues concerning the notions of authorship, publication and citation. For the sake of completeness, a panoramic report of the international scientific community that has grown around research software has been included in the section entitled A snapshot on the international RS landscape. Then, the section Key evaluation issues develops a similar analysis of the key facts concerning research software scientific evaluation, where we study the evaluation methods and the key evaluation actors, as well as the concepts of success and value of research software. Finally, the CDUR proposal section is devoted to the presentation of the proposed CDUR protocol. To facilitate the reader to reach a global perspective of this proposal, this section begins with a summary description of CDUR's four components, and then describes and studies in detail the main points of each of these components, discussing the pros and cons, and providing a used case example. The article ends with some conclusions and recommendations for the consideration of the scientific community.

---

[i]In what follows we will use often the acronym RS to refer to research software in order to facilitate the reading of this article.

[2]Our work can be also extended to deal with software driven by technological development – the development (D) in research and development (R&D), as observed in [8] (p.595) – or with scientific software being developed in private institutions or commercial enterprises, albeit requiring some specific adjustments, as, for example, the adaptation of the **R**esearch step.

[3] https://software.ac.uk/

The findings presented in this work are based on the handling of extensive literature, as well as on the first-hand, complementary experience of both authors. The first author is a research engineer at the Gaspard-Monge Computer Science laboratory (LIGM)[4], where her mission is to render RS production visible and accessible. There, she works on how to improve RS development and dissemination conditions since 2006 and has also had a similar role at national level during her participation at the PLUME project during 2007–2013 [12–15] (see the section on Publication of research software). The second author has life-long experience in research evaluation at local, national and international level in all possible scholar evaluation contexts (recruitment, career evolution, peer review, editorial, etc.). For example, he has been the mathematics coordinator for the Spanish government funding agency Agencia Estatal de Investigación[5] and, with regards to RS, he was General Chair at ISSAC in 2000[6]. On the other hand, both authors have many years of experience in RS development. For instance, the second author has recently received a distinguished software demonstration award from ISSAC 2016[7].

## 2 Research software: definition, publication and citation

In this section we examine a first block of the essential components that are involved in the scientific software evaluation, such as its definition, or more precisely, the definition of research software, and what does it mean to be an author or a contributor to this software. In the present work, we would like to highlight the widespread importance of RS production as a research output, and the relevance of the publication of software papers – as publication is an essential part of the evaluation protocol. We study as well the relationship between references and citations. Note that to cite general purpose software in scientific papers is different than to cite scientific (or more precisely, research) software specifically developed to address a concrete research problem, which is the issue here. For example, it is rather different to mention commercial software (as could be scientific software well known and widely used) in a publication than to cite the particular packages developed by a research team[8].

Under these premises, the following section entitled Key evaluation issues will be devoted to discussing some main points concerning the evaluation of RS in its own (methods, key actors, paradigms).

### 2.1 Research software definition

Generally speaking (e.g. [1]) authors consider scientific software as the one widely used in science and engineering fields. More precisely, in [16] (see also the summarized version in [17]), we can find the following definition:

*"Scientific software is defined by three characteristics: (1) it is developed to answer a scientific question; (2) it relies on the close involvement of an expert in its scientific domain; and (3) it provides data to be examined by the person who will answer that question ..."*

Or, as concisely described by [18], scientific software is *"software developed by scientists for scientists"*. Note that [16] excludes the following software types from the scientific software definition:

*"... control software whose main functioning involves the interaction with other software and hardware; user interface software that may provide the input for and report of scientific calculations; and any generalized tool that scientists may use in support of developing and executing their software, but does not of itself answer a scientific question."*

but, on the contrary, all these types could fit in the larger definition given in [1]. On the other hand, the more precise term of "research software" is also employed in the literature, a definition can be found in [19]:

*"Research software (as opposed to simply software) is software that is developed within academia and used for the purposes of research: to generate, process and analyse results. This includes a broad range of software, from highly developed packages with significant user bases to short (tens of lines of code) programs written by researchers for their own use."*

and the NASA report [10] (p.26) mentions:

*"Research software – that is, the software that researchers develop to aid their science..."*

The concept of RS is equally studied in [12] in the context of a (French) research laboratory's production:

*"Logiciel d'un laboratoire : tout programme ou fragment de programme utile pour faire avancer la recherche et qui a été produit avec la participation d'un ou plusieurs membres du laboratoire."*

*[Software of a laboratory: every program or part of a program useful to make research advance and that has been*

---

*produced with the participation of one or several members of a laboratory[c].]*

These RS definitions include some software productions that would be excluded according to the framework described in 16, such as, for example, software tools that scientists may use in support of their own developments and that could be, as well, object of research in other scientific areas.

A complete study of a lab's RS production is achieved in 12 through the comparison of software and publications, considering the legal aspects and the governance main issues in both cases. This comparison: software/publications, reconciles the different views. For instance, among scientific publications we can find preprints, articles published in conference proceedings or journals, book chapters and books. Similar diversity appears in the large spectrum that begins with research software specifically done by researchers as part of their research tasks, and that includes as well the ample concept of scientific software, widely used in science and engineering, or the notion of academic software, developed to fit education or management needs or that is developed in public institutions or in publicly funded projects.

Finally, to complete the comparison between RS and publications in the context of the study of the RS concept, let us mention that it is a general practice of research institutions and funders to evaluate the laboratories, institutes, research units... regularly. The evaluation is carried out by committees selected by the corresponding leading institutions and include the assessment of the list of publications, funded projects, PhD training, etc. Once the evaluation is over, it is also a usual practice for the publication list to become the official research output of the research unit during the evaluated period. Similarly, we think that the RS production of a research unit should be decided and proposed to the evaluation committee and, thus, it should also become part of the official research output list.

We remark that these definitions do not take into account the status of the software: project, prototype, finished, publicly available; nor its quality, scope, size, or existing documentation; it can be used by a team, just for the purpose of achieving a publication or it can be soundly installed in several labs, where it is used regularly. Moreover, we think that these considerations equally apply to software developed in any scientific area. Figure 1 shows some of the interrelations between the different concepts involved in this article.

Although we think that software development skills have improved in the last decades in the scientific community, and more and more research software developments arise from well-organized teams with well-established testing, management, documentation and dissemination procedures, the paradigmatic model that we have here in mind is one that we feel it is still largely present 12,16,17,20: software that is developed by small (perhaps international) teams, or individually, usually with few development skills, where the main goal is the research activity itself. That is, software that mainly aims to give correct scientific results of whatever kind 18,21 and not necessarily a sound software "product", well documented, tested, easily reusable or maintained. Note that *"If the software gives the wrong answer, all other qualities become irrelevant"* 16. In addition, we are well aware that part of the available RS is not disseminated adequately nor well documented, tested, correctly licensed, or maintained. Work has been done to raise awareness on these issues (see for example 13,14,22–26 among much of the cited work in this paper) in order to improve development and dissemination practices, but is the researcher's decision to balance the efforts between "software management tasks" and research efforts.

It is also a reality that much of the RS currently available is already obsolete, because of rapid software and hardware evolutions. In our vision, it is not the researchers' task to expend much time on keeping the produced RS *alive* if it is not part of their current research work. Our priority here, which is also that of Open Science, is that the code is freely and openly available, which will facilitate its examination and its adaptation to new hardware/software evolutions or to new research contexts when needed. It will also help to avoid the risk of science results based in black boxes, see for example 27 or the work of Ben Goldacre's team

---

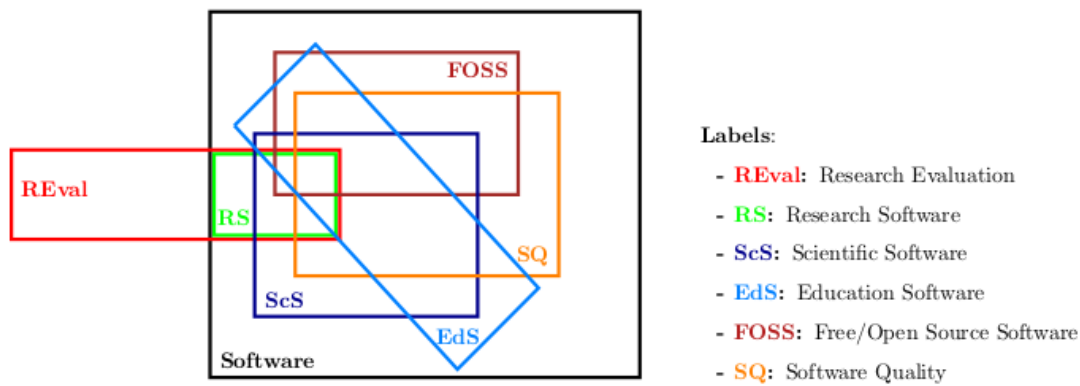

**Labels:**

- **REval:** Research Evaluation
- **RS:** Research Software
- **ScS:** Scientific Software
- **EdS:** Education Software
- **FOSS:** Free/Open Source Software
- **SQ:** Software Quality

**Figure 1. Interrelations between different software concepts appearing in this work.**

at the Evidence-Based Medicine DataLab (University of Oxford)[10]. Note that RS sustainability issues are the subject of study of institutes like SSI or URSSI as we will see in the section A snapshot on the international RS landscape below.

As we can see in detail in the section Publication of research software, there are several possibilities to publish software papers. Software publications, as it happens already for research data papers, are becoming progressively installed in some scientific areas. However, in our opinion, there is still not a real RS publication procedure of comparable status as the one achieved for research articles, that is, well established and widely adopted by the scientific community (see for example 28). Thus, we cannot rely on the concept of a research software paper to fix some features towards a precise RS definition.

Incidentally, note that *"The original idea and its implementation represent distinct kinds of contribution"* 28, and that research software can contain knowledge that is not otherwise published or just difficult to extract 29 (even from the associated publications analyzed in the section about Publication of research software) as for example: negative results, or modifications for algorithms, either to make their implementation faster and sounder or to correct problems that are arising during the coding process and that can generate new scientific understanding, etc. Code and theory exist side by side 30. An excellent example to illustrate these interactions between research and code development are found in 31 which specifically states:

> *"[...] j'aimerais aussi souligner que, dans aucune des parties de cette thèse, il ne me semble possible de séparer nettement les "mathématiques" de l'"informatique" : C'est l'implantation d'un algorithme aussi ancien que le calcul des développements de Puiseux, qui a permis d'en découvrir une nouvelle propriété mathématique."*

> *[I would like to highlight that nowhere in this dissertation seems possible to consider the "mathematics" and the "informatics" issues as something separate. It is the implementation of the old algorithm to compute Puiseux series that has allowed us to discover a new mathematical property of these series.]*

Bearing all these facts into account, we conclude that the definition of research software we will deal with in this paper must lie in the context of a research work, ongoing or already done, and, thus, in relation with a research result that is published or in construction.

Therefore, we will consider here that a research software is a well identified set of code that has been written by a (again, well identified) research team. It is software that has been built and used to produce a result published or disseminated in some article or scientific contribution. Each research software encloses a set (of files) that contains the source code and the compiled code. It can also include other elements as the documentation, specifications, use cases, a test suite, examples of input data and corresponding output data, and even preparatory material. Note the role of the preparatory material to set initial dates to the software itself (see for example the *Dates* paragraph in 12).

As remarked by a reviewer of this article, a particular RS may be considered part of a whole and may have little sense in its own, as for example the RS developed to control a specific robotic system or the distributed research software developed for Internet of Things (IoT) platforms like BeC3 (Behaviour Crowd Centric Composition)[11]. These cases are beyond the intention of this work, and the evaluation protocols proposed here are to be adapted in order to consider the whole set as a research output, and not to take into account each component separately, which depends deeply on the research community standard practices.

## 2.2 Research software authors

The concept of author is an essential part of the definition of RS: *"developed by scientists for scientists"* 18 or, simply, written by members of a research lab 12.

This is a point where legal issues are important, as the concept of author from the legal point of view might differ from the author concept in the usual scientific practice. The author of a painting is the one who holds the paintbrush, and the author of a novel is the one who writes it, which also applies to code writing. However, it is very usual that the teams involved in RS development include researchers that do not write much code, but the software will simply not exist without their scientific contribution. On the other hand, should we consider that the person who corrects a few code lines is an author? What happens if the code of a contributor has been later on fully rewritten or translated from, say, Fortran to C? In 12 we can find a complete study of the concept of RS authoring in the scientific context, as well as a study of the involved legal aspects (under the French authorship law). Note that authorship roles can also be discriminated in scientific publications, see for example 32.

As mentioned above, the role of contributors to a RS can be manifold, for example, as responsible or head of the RS developer team, as main developer or as minor contributor, as writer of an old part of the code or as the researcher supplying scientific expertise, etc. In larger teams there can be specialized roles for management, documentation writing or for testing issues. In order to simplify the analysis of the evaluation aspects implied by this concept, we have selected three main roles: (*i*) responsible or leader for the RS, (*ii*) main contributor or (*iii*) minor contributor. We should bear in mind that it may happen that RS leaders or scientific experts do not participate in writing the code, or just participate with minor contributions to the code while having other important roles in design, management, etc. They may also participate in the code writing as the main contributors. A very detailed view of several RS contribution roles can be found in 33, see also 34.

---

In conclusion, in this paper we will consider the concept of RS author as describing someone who fulfils any of the three selected roles presented above. When the contribution is about code writing, some percentage of code participation can be estimated, although some special cases can be considered. For example some percentage of code participation could be assigned to the scientific leader of the RS development team, besides the recognition of their role as coordinator or team leader.

## 2.3 A snapshot on the international RS landscape

In this section we would like to reflect the human component, in the current landscape, that is most directly concerned with research software, even if it is not always specifically related with the evaluation issues that are the object of this work. Please recall that this presentation only attempts to provide a partial view of a much larger panorama.

As the RS activities evolve in the scientific community, there is a growing organization among the RS developers and more and more initiatives of different nature build up the current RS community landscape. We can find national initiatives or institutes, national or international networks and workshops. These initiatives deal with RS as a scientific output, usually without focusing in specific scientific topics (that are out of the scope of this study). In what follows, we will mention a few examples from North America and Europe to give a glimpse of a very rapidly evolving panorama.

The first one we will like to bring here is the Research Software Engineers (RSE) initiative[12], an international association that has been launched by the UK. The last International RSE leaders meeting has taken place in January 2018[13], gathering members from Africa, America (North and South), Europe and Australia.

Again, in the UK, the Software Sustainability Institute (SSI)[14] has been launched by the EPSRC[15] in 2010[16] to help researchers build better software [22]. Let us recall that the SSI is at the origin of the UK-RSE community (see [35]) and has launched many other initiatives like the Journal of Open Science Research (JORS) publication[17], or the Fellowship Programme[18]. It also organizes many workshops regularly.

A similar structure to SSI is currently under construction in the USA, the US Research Software Sustainability Institute (URSSI)[19] that aims to develop a pathway to research software sustainability.

From a different perspective, the Workshop on Sustainable Software for Science: Practice and Experiences (WSSSPE)[20] is an international community-driven organization that promotes sustainable research software by addressing challenges related to its full lifecycle, through shared learning and community action. It has organized workshops in USA and Europe since 2013.

The Software Citation Group has published the Software citation principles in 2016 [7]. The group is now closed and has evolved to the Software Citation Implementation Working Group[21]. They are specialized communities, focusing on the particular issue of software citation.

In France, the PLUME project (2006–2013)[22] was launched by the CNRS[23] to promote economical, useful and maintained software for the higher education and the research community, and to promote the community's own developments [13]. PLUME had a platform to publish software descriptions and reference cards (*fiches* in French) and organized many workshops and training activities around research software, see for example ENVOL 2008[24] or the *Journées PLUME 22-23/09/09 : Pourquoi et comment diffuser un développement logiciel de laboratoire ou d'université en libre?*[25]. PLUME has also published several studies regarding RS issues that can be found under the title *Patrimoine logiciel d'un laboratoire*[26].

In France there is also the developers' network named *Réseau des acteurs du DÉVeloppement LOGiciel au sein de l'Enseignement Supérieur et de la Recherche* (DevLOG)[27] that was launched in 2011 to gather the actors of the software development in the academic community. It has regularly organised the JDEV conference since 2011[28].

Canarie[29] is Canada's National Research and Education Network. It is a funding agency with a Research Software Program[30]

that has organized the Canadian Research Software Conference in 2018[31].

In Netherlands, the NL-RSE (Netherlands Research Software Engineer community)[32] was formed in April 2017, an initiative launched by the Netherlands eScience Center[33] and ePLAN[34] and had its first meeting in 2017. In Germany, the deRSE[35] organized its first Conference for Research Software Engineers in 2019[36].

Besides these initiatives, we can mention many different conferences or workshops around research or academic software, such as the Engineering Academic Software (Dagstuhl Perspectives Workshop 16252)[37] (June 2016) 23 or the DANS[38] /SSI-Sustainable Software Sustainability Workshop[39] (March 2017) with a follow up in 2019[40].

We consider that these references provide a relevant, albeit partial, snapshot of a situation evolving towards an increasingly internationalized structuration[41].

## 2.4 Publication of research software
In this section we will present a partial panorama of the RS publication world. *Research software papers*, like research data papers, are publications directly related with the description and functioning of the RS, published in a scientific area's specific journal or in generic journals.

In order to give a glimpse of the current panorama for these publications, we can begin with the list that N. Chue Hong keeps in the SSI platform[42]. Among the general scope journals mentioned in N. Chue Hong's list, Wiley's *Journal of Software: Practice and Experience*[43] has published articles about software in science since 1979[44] and seems to be one of the oldest journals for this subject. On the other hand, the Research Ideas and Outcomes (RIO) Journal[45] has published, at the time of writing this paper, three software descriptions and one software management plan, and seems still too novel concerning this kind of publication. The recent Software Impacts[46] journal two first volumes are in progress in October 2019.

In a similar mood regarding RIO's RS descriptions, we can mention the software publications by *The Journal of Open Source Software (JOSS)*[47] 36, a "developer friendly" journal launched in 2016 that has review policies taking into account good dissemination practices (see for example 14,24). Reviewers are expected to install the software they are reviewing and to verify the core functionality of the software[48].

Another journal considering RS submissions is the *Elsevier SoftwareX* Journal[49], launched in 2015, where the peer review procedure requests the referees, among others tasks, to build, deploy, install and run the software following the provided documentation[50]. Similarly, JORS[51] has published *software meta-papers* since 2013 and also has a precise *software peer review* procedure[52].

In France, the PLUME project (2006–2013) 13,15 published software papers of several kinds. We only describe here briefly two: RS publications and validated software publications. We describe first the RS description cards[53] (also named reference cards or *dév Ens Sup - Recherche* cards or *fiches* in French) that are short publications containing the software metadata, a brief description, and links to related publications (see for example the Treecloud[54] reference card in Figure 2). Similar descriptions can be also found at the Netherlands eScience Center[55].

The objective of these publications was to promote and increase the visibility of RS, in order to favor scientific collaboration[56]:

> *"Ces fiches sont destinées à valoriser les productions logicielles de la communauté ESR et à les faire connaître auprès d'un public plus large dans un but d'augmenter les collaborations scientifiques."*
>
> *[These description cards are intended to promote the software production of the research community and to make it known to a larger public in order to increase scientific collaboration.]*

The other kind of software publications described here were dedicated to software with the status "*validated by the academic*

---

[31] https://www.canarie.ca/software/canadian-research-software-conference/canadian-research-software-conference-2018/program/

[32] http://nl-rse.org/pages/community.html

[33] http://www.esciencecenter.nl/

[34] https://escience-platform.nl/

[35] https://www.de-rse.org/en/

[36] https://www.de-rse.org/en/conf2019/index.html

[37] https://www.dagstuhl.de/de/programm/kalender/semhp/?semnr=16252

[38] DANS is the the Netherlands institute for permanent access to digital research resources, see https://dans.knaw.nl/en

[39] https://dans.knaw.nl/nl/actueel/software-sustainability-workshop-7-9-march

[40] https://software.ac.uk/wosss19/agenda

[41] https://researchsoftware.org/

[42] https://www.software.ac.uk/resources/guides/which-journals-should-i-publish-my-software

[43] https://onlinelibrary.wiley.com/journal/1097024x

[44] Unfortunately, we have no free access to this journal, which hinders our goal of presenting a short comparison with more recent approaches to RS publishing.

[45] https://riojournal.com/

[46] https://www.journals.elsevier.com/software-impacts/

[47] http://joss.theoj.org/

[48] https://joss.readthedocs.io/en/latest/review_criteria.html

[49] https://www.journals.elsevier.com/softwarex/

[50] https://www.elsevier.com/__data/assets/pdf_file/0010/97066/ReviewerForm.pdf

[51] https://openresearchsoftware.metajnl.com/

[52] https://openresearchsoftware.metajnl.com/about/editorialpolicies/

[53] https://www.projet-plume.org/fiches_dev_ESR, http://www.projet-plume.org/en

[54] https://projet-plume.org/en/relier/treecloud

[55] https://research-software.nl/

[56] https://www.projet-plume.org/types-de-fiches#dev_ens_sup_recherche

**Figure 2. Initial section of the TreeCloud reference card published in the PLUME project platform, see https://www.projet-plume.org/en/relier/treecloud.**

*community in the sense of PLUME"*. The concept of *validated software* is a tricky one. Between the software developed by a little team and a software widely known, adopted and used by a large community, there is a whole range of possibilities. Where should we put the limit in which a software can be declared validated? The choice of PLUME software was: if it is known to be used regularly in at least three laboratories or institutions[57]. This translates the question of how to declare a software to be validated by identified experts to the issue of finding how to get the knowledge about its regular use in at least three academic institutions. In the case of PLUME, the team contacted (or was approached by) experts that were willing to write validated software descriptions and share their knowledge about the software. We can emphasize that, although PLUME did not have a software peer review procedure, the validated software publications where reviewed and completed by identified users who also had first-hand knowledge about the software (see 15 for a short description of the validated software publication procedure).

The following statistics can be found in the PLUME platform: 406 validated software cards, 358 RS French reference cards and 116 English RS reference cards where published before the end of 2013 with the collaboration of 950 contributors.

Among the 406 validated software cards, 96 correspond to research software[58]. This means that some RS had associated three different *fiches*, which increased its visibility. See for example the three publications associated to the Unitex software developed at the LIGM laboratory: the Unitex English reference card[59] is a translation of the French card[60] and both indicate that there is a more detailed description card as a validated software[61] (see Figure 3).

---

[57]https://www.projet-plume.org/types-de-fiches#logiciel_valide

[58]Validated software descriptions are only available in the French side of the platform. Part of these validated software were developed by researchers, see https://www.projet-plume.org/fiches_logiciels_dev_internes for the list of RS validated software descriptions published in French.

The PLUME software publications are classified by subject (maths, network, biology...) and by keyword tagging. For example, a set of institutional keywords was used to identify the *dév Ens Sup - Recherche* cards produced with the joint participation of members of a laboratory or a research establishment. Both approaches (subject classification, keywords) facilitate software searching, as it is very difficult in general to find a specific RS if one does not know its name or its developer's team. For instance, the difficulties of finding software of interest for scientists and engineers have been thoroughly studied in 4,5, yielding relevant inconveniences such as, for example, the risk of work duplication, of reduced scientific reproducibility and of poor return of funding agencies' investment. Of course, the most serious drawback is the real reduction of the RS's potential scientific impact.

Although it does not fall in the category of general journals, we would also like to mention the Image Processing On Line Journal (IPOL)[62] (see 37), an *Open Science and Reproducible Research journal*. IPOL publishes image processing and image analysis articles, and the source code associated to the described algorithms. Both the submitted article and the corresponding software are reviewed together and published simultaneously in the IPOL platform. The software can be tested to verify the correctness of the published results or to generate new knowledge with other data.

Finally, as a general reflection regarding peer review procedures specifically suited for research software articles, we would like to highlight the "Peer Community in"[63], a non-profit scientific organization that aims to create specific communities of researchers reviewing and recommending, for free, unpublished

---

**Figure 3. Initial section of the Unitex validated software description card published in the French side of the PLUME project platform, see the complete publication on https://projet-plume.org/fiche/unitex.**

preprints in a scientific area. This initiative seems to us an interesting experience that could be well adapted to develop RS recommending and review. It also seems particularly well adapted to declare some RS as *validated* by a well identified community.

Comparative software reviews are other forms of reviewing that could be useful to improve RS assessment inside scientific communities or for specific targeted topics, see 38 for counsel in writing such reviews.

## 2.5 Referencing and citation of research software
Before the emergence of a specific approach to software papers publication, as described in the previous section about Publication of research software, scientists used to present research software through the usual scientific article media. For example, we can find in 28:

> *"Academic computer science has an odd relationship with software: Publishing papers about software is considered a distinctly stronger contribution than publishing the software. The historical reasons for this paradox no longer apply, but their legacy remains."*
>
> *[...] Typically, scholars who develop software do not include it as a primary contribution for performance reviews. Instead, they write articles about the software and present the articles as contributions."*

Many of these traditional articles did (and still do) present scientific software to some extent, from a light mention of the involved software within a more or less detailed description of the core scientific work, to a very thorough explanation of the RS, including its name, version and retrieval indications. Thus, quite a few of these papers would have been considered today as RS papers, as it might have happened that the corresponding reviewers have had the opportunity to carefully check the involved RS or, at least, they have had some first-hand knowledge about it (as users from the corresponding research community, for example). That is, the reviewers may actually have behaved, albeit informally, as true RS reviewers, meeting current requirements of journals like JOSS or JORS (see previous section).

What is, yet, less usual is to set a reference to one of these papers as *the* reference for the RS being described in the article. And, even when a reference is treated in this way, it is still rare that it assigns, to the corresponding RS, the status of a research output on its own, since the usual situation is to understand that the most valuable contribution is the article itself, as seen in 28. That is, the researcher, and its evaluators, still consider that the value remains in having another article added to the curriculum vitæ publication list, mostly forgetting the relevance of adding a RS to the list of the research production outcomes, or to take it into account in the evaluation process. This is a serious drawback, both towards standardizing RS citation procedures as well as for its full recognition as a research output, which is also remarked in 28: *"This conflation of articles and software serves neither medium well"*, but it is nevertheless accepted in this paper as a current practice.

Nowadays, as the community value perception switches towards including RS, research data and other outputs as fully recognized elements of research production, the issue of how to properly cite these works sharply raises. Many scientific groups currently deal with this question, mainly for research data. For the case of research software we would like to mention, for example, the Software Citation Group 7, the Software Citation Implementation Working Group[64] and the Dagstuhl Perspectives Workshop 16252: Engineering Academic Software 23 and also the RELIER Group[65], the RS team of the PLUME project 15.

In general, we consider that the first step required to facilitate RS citation is to have associated, from the origin, each RS with a clear reference. As it is explained in 39:

> *"... la différence entre référence et citation : l'acte de référence relève d'un auteur donné alors que la citation est une nouvelle propriété, éventuellement calculable, du texte source. Selon P. Wouters (1999), ce renversement a radicalement modifié les pratiques de référencement et littéralement créé une nouvelle "culture de la citation"."*
>
> *[… the difference between reference and citation: the act of reference is the responsibility of a given author while the citation is a new property, possibly calculable, of the source text. According to P. Wouters (1999), this reversal has radically altered the practice of referral and has literally created a new "culture of citation".]*

That is, to establish a precise reference is the authors' task, so that the interested public can use this reference to cite the corresponding work. As stated in the above quotation, reference usage is radically evolving and is creating a new citation culture.

In this paper we will consider *a RS reference* or *citation form* as (see 40, section 6.1):

- the reference to a research software paper or other kind of scientific publication that includes, and relies on, a software peer review procedure, or

- the reference to a standard research article that includes a description of the RS and the implemented algorithms, explaining motivations, goals and results, or

- a typical label, associated to the RS itself, and that identifies it as a research output, specifying its title, authors, version, date, and the place the software can be recovered from. In this respect it can be relevant to make the code citable[66] via repositories such as Zenodo 41.

Note that a RS can have simultaneously more than one of these types of reference forms, and, in fact, all three types can coexist for the same RS. Moreover, concerning the second reference type, we remark that there can be several classic articles associated to a single RS, and that could be used as its reference, see for example many of the RS related articles published by the

---

[64] https://www.force11.org/group/software-citation-implementation-working-group

[65] https://www.projet-plume.org/relier

[66] https://integrity.mit.edu/handbook/citing-your-sources/avoiding-plagiarism-cite-your-source

Journal of Symbolic Computation[67] or the articles listed in the PLUME RS descriptions[68]. Thus, in order to facilitate RS citation by others, it is advisable to choose the simplest reference form, whenever this is feasible. This can be easy for a small RS that has been produced and used for a unique purpose and described in just one research article and that will not evolve any further. But RS can be a living object with many evolutions, versions and different references associated to it, making it harder to decide on a single citation form.

A further issue concerning RS citation arises from the observation that, in the case of research papers, the classic publication model attaches only one reference and only one reference to each paper. Thus, the paper and its reference act also as a timestamp for a published scientific result. This model is still largely set in place, but it is evolving and now journals like F1000 Research[69] manage quite a few versions of the same article as part of its open peer review practice [42]. Furthermore, in [43] the sixth key principle claims that *"the idea of the journal article as a monolithic object that will stand for all time unless formally retracted has gone"*. Indeed, unlike articles, software development has never been associated to a monolithic object, and there exists RS with a long life, involving several international collaborations, with a large number of versions and of related publications that could act as a reference. Moreover, some RS users would prefer to cite a specific version of the RS, as the one that has been included in other RS, or that has been used to obtain the published result.

The use of persistent identifiers[70] such as DOIs facilitate RS access and it is advisable to include these identifiers in the citation formula.

A more complex way for RS identification than a citation form is the use of metadatasets. The Software Citation Implementation Working Group is working over several possibilities for software metadatasets[71]. The PRESOFT (Preservation for REsearch SOFTware) metadataset proposed for RS in [40] is built over the skeleton of the RS reference cards that where published between 2008 and 2013 by the PLUME project [15]. This metadataset benefits from the PLUME experience, which validates the proposed model, and sets a reasonable level of standards to manage RS identification.

Further discussions on software reference and citation issues can be found in [3,4,6,15,34,44–46] and a thorough digression on software metadata appears in [47] section 6. We can also mention that [48] addresses a related topic, namely, the proposition of a usage format for CITATION files.

These works, among many others, as well as the contributions of different software and data discussion groups, and the previous reflections in this section show the complexity of the concept of RS citation and its actual evolution towards a more well-established future model(s).

Finally, let us recall the growing importance of citations, DOIs and other tools to build Open Science Graphs, a subject recently explored in the OS FAIR[72]. In particular the work [49] explores Software and their dependencies in Research Citation Graphs.

## 3 Key evaluation issues

This section outlines the second block of issues that appear in our conception of the CDUR procedure for RS evaluation, as are the possible evaluation situations, the evaluation methods, the main evaluation actors and the study of the concept of a successful RS and its comparative value regarding other kind of research contributions.

### 3.1 Evaluation situations

If we follow the academic life of a standard researcher, we can appreciate that research evaluation starts with the doctoral thesis, followed by the recruitment period and continues, along the years, with the career evolution. Subject to this research evaluation are, in particular:

– articles and other publications with peer review procedures,

– participation at congress and workshops requiring invitations or with refereed submission procedures,

– applications at different competitive calls for project funding, proposal and/or involvement in different research projects (perhaps with the collaboration of other colleagues, institutions or technological enterprises...),

– as well as the establishment and consolidation of a network of students and of reputed colleagues, usually in an international context and within a scientific area.

Collaborations tend to be more and more interdisciplinary, which raises difficulties in the evaluation process, as evaluators can be experts in one area but have little knowledge in the other involved areas. As remarked in [50], there is also a lack of widely agreed quality principles in interdisciplinary research, and reviewers might lack cross-disciplinary experience and, maybe, do not have shared understanding of quality. Interdisciplinary aspects can appear in RS evaluation, as software conceived and developed in a specific research area can be easily used and adopted (including with new developments) in many other areas. This aspect is part of the usual research evaluation considerations.

Let us remark that the first evaluation that usually comes into play is the one realised by the researcher in its own, as it is the researcher that first considers if a result, a paper, a RS is ready to be disseminated in a precise context, ready to be sent to a journal, or if an idea of a project has any chance in a funding call. For

---

example, a RS can be first disseminated with the goal to look for collaborations, without giving any importance to its early development status. It is the researcher who decides to disseminate a result on a preprint deposited on arXiv[73], or to send it for a more sound publishing procedure in a scientific journal. It is also the researcher who first evaluates when some work is ready for external examination in one or other manner. Likewise, the researcher decides if a preprint deposited in arXiv will be the object of future work, or be forgotten forever, and similarly for RS, the researcher decides which will be the effort to put in the software management tasks (documentation, testing, versioning and maintenance). These decisions usually can find variations following evolution in research interests, funding constraints and fluctuate when facing evaluation issues such as recruitment.

On the other hand, as it is noted by a reviewer of this article, it is also usual to work on several year projects, thesis, postdoctoral positions and other temporary chairs. In this context, the evaluation that comes into play is the final project evaluation, in the case of funded projects, or the thesis defence. For other temporary positions, it is the team evaluation that may come into play. For example, if the work involves some RS development, the team should verify if the RS works correctly and have enough documentation and testing procedures in order to facilitate the possible follow ups. This is to be done before the developer leaves the team. Thus, even if there is not a real dissemination, some kind of internal dissemination and examination procedure should be considered, which applies to RS as well as other outputs, in order not to hinder the future teamwork, avoid waste of time and to maximize mutualisation.

Any dissemination procedure has its own goals. It also has a target public, even if it is a very restricted one: the collaborative team, the thesis jury, the funded project evaluators, funders, etc., and the researcher prepares their work in adaptation to the related evaluation or scrutiny surely raised by the dissemination context.

### 3.2 Two evaluation methods
Generally speaking, we can identify two main methods to conduct research evaluation, both of which usually take into account only the "paper" production (articles, books, presentations, project proposals, etc.): a qualitative approach (sit and read to evaluate, following a subjective criterion, the quality of the presented documents) and a quantitative estimation, by using established external metrics such as the citation index, impact factor, number of publications and other indices, and then adopting some models of evaluation 8. Nevertheless it is widely known that indicators and bibliometrics should be used with careful attention, as remarked in 11,39,51–54.

Moreover, the community's "social knowledge" can also occur and influence evaluation practice, as stated by 55 (p.8) regarding the review of mathematical statements in research papers: *"the methods of checking the proof are social rather than formal"*. That is, evaluation of the quality of a work can rely, although maybe not intentionally, upon a community perceived knowledge.

The extent of these social practices is very difficult to assess but could be neutralized by increasing openness and transparency in the evaluation policies and by exercising special caution during the evaluation committee selection, as recommended by 11 (p.9):

*"Universities and research institutions should:*

*4. Strive for a balanced and diverse representation (including, but not limited to, gender, geography and career stage) when hiring, seeking collaborations, when organizing conferences, when convening committees, and when assigning editors and peer-reviewers, and building communities such as learned societies.*

*[...] Research funders and policy-makers should:*

*2. When evaluating researchers, ensure that a wide range of contributions (scholarly publications, but also data, software, materials, etc.) and activities (mentoring, teaching, reviewing, etc.) are considered, and that processes and criteria of evaluation are both appropriate to the funder's research programme, and transparent."*

### 3.3 Key evaluation actors
As we have seen in the previous section, there are, all along a research career, different key actors performing the evaluation tasks at different stages: for example, the research community as a whole, through its experts, regarding peer review procedures for publications and journal editorial activities; the academic community of colleagues from universities, laboratories or research units, involved in evaluation processes for recruitment and career progress; the committees nominated by funders of scientific activities at local, national and international level; and, finally, the policy makers at any level, that set the policies that will be applied by the selection, recruitment or the funding committees.

On the other hand, community evaluation appears while setting new collaborations or in the gathering of a team looking for some project funding. Besides, informal evaluation also happens each time the reader of an article or a RS user weighs if further attention to the research object is worth it: is this paper or software interesting for my work/research?

A researcher aiming to achieve success during any evaluation of whatever kind needs, first to submit good research outputs (articles, software…) and, then, to make public this work adequately in order to facilitate the evaluator's task. But it is not the same to face a journal peer review procedure, a grant selection for project funding, or to be involved in a recruitment process, etc. Similarly, it is not the same to be subject to an evaluation by qualitative or quantitative methods. As a consequence, consciously or not, author's adaptations occur when facing evaluation procedures and requirements 52.

On the side of policy makers, it is necessary to foresee and to adapt to science evolutions, and these challenges also ask for new evaluation policies and criteria. We would like to mention here three examples of the preparation of such policies. The first one corresponds to the Expert group of the European Commission that has produced the report 11. This Expert group has been set up to support the policy development of the European Commission on Open Science, in order to assess the current situation with regard to scholarly communication and publishing mechanisms, as well as to propose the adoption of some new, general, principles for the future. The second example corresponds to the Committee on Best Practices for a Future Open Code Policy for NASA Space Science that has produced the report 10. This Committee was charged to investigate and recommend best practices for the NASA, as it is under study whether to establish an open code and open model policy, complementary to its current open data policy. Both of these two reports do provide recommendations regarding evaluation, RS and Open Science (as we include open code in the general framework of Open Science policies). The third example that we would like to mention is slightly different, as it corresponds to the Symposium organised by the French *Académie des sciences* (April 2nd 2019) for *Foresighting Open Science*[74]. Among others, the goal of the symposium was to look into the issues that the current Open Science acceleration raises, such as science and researchers' evaluation. The *Académie* acts as an expert and advisory body for the French public authorities. These examples show us how the policy makers set expert committees or organize events to study a particular subject and to seek counsel for the new policies to be defined.

Another important role of the policy makers is to set the evaluation committees, as there is a fundamental distinction between who establishes the norms, policies or habits in the evaluation procedures, and the evaluator or the evaluation's committee, who has to apply them. Yet, it can also happen that the roles of the policy maker and the evaluator are concentrated in the same person or persons. Policy makers set not only the evaluation methods to be applied, but also the characteristics and criteria of the jury's selection and whether the committees are totally independent and have the final decision or are just an advisory board, etc.

In particular, issues such as gender, age, race, geography, etc. biases can be better dealt with through committees with a balanced representation of diversity 11 (p.9):

> *"Research funders and policy-makers should:*
>
> ***4.*** *Consider how funding policies affect diversity and inclusivity of research on a global scale. In particular, funders should work to ensure that review boards, committees, panels, etc., are diverse - in terms of gender, geography, and career stage."*

Further considerations on the evaluation role of universities, scientific establishments, funders and policy makers are addressed in the next section dedicated to the CDUR protocol.

Finally, let us to point out that when evaluating a publication or when performing peer review of an article, the evaluator is expected to have the necessary knowledge to recommend if the document should be published. At the end of the review process, the evaluator is expected to have a fair amount of knowledge about the reviewed work. Similarly, in a recruitment procedure, the evaluator is expected to have the necessary knowledge in order to decide the best candidate for the position. But these arguments are not obvious concerning RS evaluation, as we will detail in the CDUR proposal section.

### 3.4 Towards a successful research software

Obviously, a good scientific software is one that has been written by a scientific team to produce good, sound scientific results. This is quite a circular definition, and other, more precise, criteria should be taken into account. For instance we could consider as a positive feature the RS availability, and the fact that it is adequately disseminated, documented, licensed, version controlled, and tested 19,25,28. In 56, the proposed criteria to assess RS are Availability, Usability, Maintainability and Portability. Other qualifying principles that are currently under discussion are the Software Seal of Approval or those involved in the notion of FAIR (Findable, Accessible, Interoperable and Re-usable) software, as FAIR is already a very popular concept among the research data communities 57.

What is less obvious is to determine how these criteria will be concretely used in a specific evaluation context. In fact, setting the list and the weight of the different criteria to determine what should be understood as a good RS depends on three different aspects:

- the evaluation context (peer review, funding, career...),
- the evaluation committee,
- the policy makers.

To continue the study of the criteria that could be considered to declare a RS as successful, we can recall that the scientific community has clearly established the difference between an article and a preprint through the peer review process that is part of the publication step and that is missing in the preprint case. As a result, an article has a quality label that does not exist in the preprint case.

Now, in the same way as preprints are made publicly available through their deposit in arXiv or in many other popular platforms, software can be also made publicly available in platforms like Zenodo[75], GitHub or many others. Although there may be some control of what is deposited in these platforms, there is not (as far as we know) a *real scientific control* of the deposits, or something that approaches peer review procedures.

Again, a distinction similar to the one existing between a preprint and a published article can be claimed for software: there

---

is a clear difference between RS publicly available through well known platforms (or personal web pages) and RS that has been the object of a publication of the kind detailed in the section of Publication of research software. As we have seen there, RS reviewers are generally expected to have both sound scientific knowledge and enough software knowledge to be able to build, deploy, install and run the software following the provided documentation. This is, then, a dual context where the evaluation of purely software aspects has to go in parallel or, perhaps, get mixed with the evaluation of scientific aspects.

This level of evaluation of software aspects can be adequate for a RS paper, but it could be less adapted to recruitment or a career evaluation process, where evaluators have got to achieve a global vision of a curriculum vitæ.

Another relevant remark is related to the assessment of those RS products which are already well known and popular within a scientific community. Here, it could be more adequate to assess the RS quality or its impact in some indirect way, by assessing the quality of the related publications, or by the number of users and collaborators that have been attracted, by the number of funded projects, etc. (12, p.134). Yet, this quality test is to be carefully considered, as a RS considered as successful from the software point of view is not necessarily good from the scientific perspective, and vice versa (12, p.134).

In 26, a RS is considered as successful when it is delivered with a base code that produces consistent, reproducible results, when it is usable and useful, when it can be easily maintained and updated, and has a reasonable shelf life. The French Inria institute uses a set of criteria for software "self-assessment" in career evaluations in order to determine software quality 34,58. On the other hand, as seen in section about Publication of research software, PLUME handles the concept of validated software, based in the verification of its regular use in at least three different laboratories (avoiding in this case the need for a careful analysis of the code). Going beyond PLUME's concept of validated software, a very successful RS could be simply defined as one that is widely adopted and used by a scientific community.

Whether we are evaluating a RS or a contribution to a RS, the RS needs to be well identified with a version number and a date, that is, with a reference of the kind proposed in the section about Referencing and citation of research software. The role of the different agents contributing to the RS should also be clearly presented, as seen in the section about RS authors.

Finally, concerning the role of policy makers and of evaluation committees in setting and applying the different criteria used in RS evaluation, we would like to emphasize that they should clearly state the precise balance between the scientific and the purely technical aspects that has been considered in the evaluation process, as this is a key consideration to understand the concept of successful RS that is behind the evaluation outcome.

Further considerations on this manifold notion are the object of the CDUR procedure proposed in the next section.

## 3.5 The comparative value of research software
Once we have analyzed the different criteria to determine when a given RS could be considered as good or successful, an important issue that remains to be settled is how this success should be reflected in the more general value scale of a research evaluation process. To our knowledge, the question of the comparative weight between a relevant RS and a good research article arises repeatedly: should the evaluators assign to the RS the same value as to the article?

In this context, it seems relevant to analyze a similar situation that happens in the well-established publication evaluation scheme[76] when considering different publication outputs: preprint, conference proceedings, journal published article, book chapter, book... It seems to us that, in this case, the value scale for the different products is widely accepted in whatever scientific community, although idiosyncratic variations can apply (for example, to manage a blog can be taken into consideration in Humanities and Social Sciences evaluations, but not in Mathematics). For instance, journal published articles are usually considered better if there is a rigorous peer review procedure behind the acceptance procedure, but, again, the specific scientific ecosystem determines the criteria to declare which journals are better than others (leaving aside, voluntarily, the arguable journal impact factor consideration).

In a similar way, we could tentatively try to set a RS value scale, backed by the standard uses of a specific community. Say, proposing a scale that would start with a RS with only one associated article, followed by a RS with several associated articles by the same team that has produced the RS, and then a RS that is used by other teams to produce new published results, a RS that has passed a strict software peer review (as in the section Publication of research software), and finally, up to a RS being the object of international collaborations during months or years... We think that a way to solve this problem of comparing the value of publications and RS is to give each of them the value that they have achieved in the corresponding, specific, scale. Likewise, for other research outputs (data, prototypes...) that are not publications or RS, they need to have their own scale of value too. The policy makers should build the corresponding scales and explain how they will be applied, mainly in comparison to the publication scale, while respecting the traditions and functioning of the involved research community or academic institution.

## 4 The CDUR proposal
In this section we detail our proposal for the evaluation of research software. We have labelled our proposal with the acronym CDUR, that stands for **C**itation, **D**issemination, **U**se and **R**esearch. In what follows we will introduce each of these items from multiple perspectives, referring to the policy makers, to the evaluators and to the evaluated researchers, that is, to the key evaluation actors, as seen in section about the Key evaluation actors. Moreover, we have chosen to present our proposal, first, in a summarized way, followed, then, by an extended description that develops in detail a multiplicity of choices for a very flexible application of the CDUR protocol. This extended description is followed by a practical use case example and some final considerations.

---

[76]For a recent analysis of different behaviors regarding publication practices and Open Science issues, you can see the (French) presentation of Jean-Pierre Bourguignon, President of the European Research Council, the 2nd April 2019, at the Académie de Sciences in Paris, https://public.weconext.eu/academie-sciences/2019-04-02/video_id_010/index.html

### 4.1 The CDUR procedure: a summary
The CDUR protocol contains four steps to be carried out in the evaluation of a RS. These steps are to be applied in the following chronological order: **C**itation, **D**issemination, **U**se and **R**esearch. For example, we consider that, to facilitate dissemination, a RS should be, first, a well identified object; and in order to be correctly cited, the RS reference should be clearly indicated, as argued in the section about Referencing and citation. Let us introduce a resumed version of these four steps.

(**C**) **Citation**. This step measures to what extent the evaluated RS is well identified as a research output, as a research object in its own. It is also the step where RS authors are correctly identified as well. We have seen in the RS publication section three different ways to establish a RS reference, in order to facilitate its citation. Moreover, a more evolved RS identification level could be provided in the form of a metadataset. Reference and metadata include, among other information, the list of the RS authors and their affiliations, as seen in the section about RS authors.

(**D**) **Dissemination**. This step measures the quality of the dissemination plan for the RS, involving actions such as (see 14,24):
- Choosing a license, with the agreement of all the rightholders and authors. Consider, preferably, using free/open source software licenses.
- Choosing a web site, forge, or deposit to distribute the product; stating clearly licensing and conditions of use, copy, modification, and/or redistribution.
- Creating and indicating a contact address.

This is the step related to legal issues dealing with the authors and rightholders (as established in the **C**itation step) deciding and installing the license(s) for the RS dissemination 12,14,59,60. This is also the step concerning Open Science, as the RS license expresses its sharing conditions; and the step where policy makers should establish the Open Science policies that will be applied in the evaluation process.

Finally, let us recall that the inclusion of the list of related publications, data sets and other related works in the dissemination procedure helps to prepare the reproducible science issues that are to be taken into account in the **U**se step.

(**U**) **Use**. This step is devoted to the evaluation of the technical software aspects. In particular, this step measures the quality of the RS usage, considering that a performing RS is one that is both correct and usable by the target scientific community.

The RS usability does not only refer to the quality of the scientific output but also can deal with other matters, such as the provided documentation, tutorials and examples (including both inputs and outputs), an easy and intuitive manipulation, testing and version management, etc. 25.

This is the reproducible science step, where the published results obtained with the RS should be replicated and reproduced 37,61–63.

(**R**) **Research**. This step measures the impact of the scientific research that has required in an essential way the RS under consideration.

The evaluation of this item should follow whatever standards for scientific research quality in the concerned community (e.g. 8,52,64,65).

This is the step where the RS related publications (as described in the RS definition section) come into play, and where the evaluation should consider the difficulty of the addressed scientific problems, the quality of the obtained results, the efficiency of the proposed algorithms, etc. The RS impact can also be assessed through the research impact of the related publications, and through its inclusion (or use) as software component in other RS.

Finally, the CDUR procedure is meant to assist the evaluated researchers, the evaluator committees and the evaluation policy makers. It can be easily adapted to many different RS evaluation situations. It applies equally to any scientific area, as we concentrate our evaluation protocol in the general RS aspects, concentrating in the **R**esearch step those aspects specifically related to some particular areas.

### 4.2 The CDUR protocol in detail
As we have seen in the previous summary, the CDUR's RS evaluation procedure considers a RS as a scientific output, measures the way it is identified and disseminated, and then takes into account the specific software and research aspects. We should remark that the realization of some Software Management Plan (SMP), such as those proposed in 40,66, can help the development team to reflect on and to prepare the different evaluation points. These plans can be made public and released jointly with the RS. In this way, SMPs can also help evaluators to achieve a better RS assessment. Note that the provision of such SMPs can be part of the evaluation process policies, as it is already the case of Data Management Plans when applying for EC funded projects 67.

Regarding precedents to the CDUR protocol we can mention the Inria software description form 68, that was proposed in 2007 for RS assessment, where we can find points in common with the CDUR protocol. Both, the Inria form and CDUR have common ground with the software peer review methods mentioned in the RS publication section. In fact, to fill forms like the one proposed at Inria can help in the preparation of the CDUR evaluation. And, what is more relevant, as it happens with the SMPs, such Inria forms might help yielding and guiding some key reflection issues over the evaluated RS that could be, otherwise, somehow forgotten by the RS development teams. The generalized use of these or other similar forms should be decided by the RS evaluation committees or by the policy makers to set some standards in the evaluation procedure.

In what follows we will analyze and develop in detail the different points of the CDUR protocol.

**(C) Citation.** In order to facilitate the RS citation, authors must set a reference. As we have seen in the Referencing and citation section, there are at least three possibilities for RS referencing and, moreover, they can cohabit. In order to facilitate the citation by others, one reference form should be put forward.

The reference should indicate, among others, the RS authors and their affiliations. In the case of a large list of contributors, it should give the list of the main contributors and refer to other documents (web page, RS documentation...) for the complete list of contributors. This basic step could be completed with the inclusion of DOIs or other persistent identifiers in the reference form.

A more complex way for RS identification is the use of metadata sets. Note that the existing RS metadata sets can be adapted or completed in order to fit many different evaluation situations.

We consider that it is the RS authors' role to set the best way to cite or identify their RS. On the other hand, and following the software citation principles in 7 (see also 23), authors should cite correctly other related works, with thorough attention to cite other research software components that could have been included in the RS or on which the RS depends upon.

Citation and metadata are the tools to measure how easy is the access to RS. It is the role of the policy makers to set the required citation or metadata level that is best adapted to the evaluation context at stake, either by fixing a reference format or with adapted metadata sets.

Finally, let us recall that evaluators should verify that the RS under evaluation complies with the citation or metadata characteristics required in the evaluation, and they should also check the correctness of the citations, in the given RS, to other external RS works.

**(D) Dissemination.** RS dissemination should take into account that it can target different levels and types of public: from a very restricted set of persons, such as the closest collaborators of the RS team or the evaluation committee itself, up to the most general collection of addressees, through the dissemination of the RS via a web page, a software forge or deposit, either oriented to a scientific area or to a very general public. Moreover, in the case of a restricted dissemination of a RS having a widely available reference (because of the existence of related publications, for example), it is advisable to include a RS mail contact address to facilitate potential scientific collaborations.

In any dissemination procedure (as the ones proposed in 14,24) the RS sharing conditions should be clearly stated, as the running, loading, reproducing, translating or arranging of a computer program can only be done upon the corresponding (written) authorization, as stated by the law[77]. Thus, RS dissemination documents should include a license[78] (or a written agreement, in the case of a restricted public dissemination) establishing the sharing conditions and describing the legal framework where the RS can be used, compiled, reproduced, copied etc. 12,14,59,60. Only the rightholders can decide and set the RS license, hence the importance of including the list of authors and its affiliations, as described in the **C**itation step. Note that the license information could be included in the citation formula, and it is usually included in the RS metadata (using for example standards like SPDX (Software Package Data Exchange)[79].

The license will determine if the software is free[80] and open source[81], and whether, as a research output, its dissemination fits the Budapest Open Access Initiative guidelines[82]. In this direction, the report 9 considers as a positive criterion in the evaluation of research careers, the regular use and development of free/open source software, fully acknowledging Open Science practices.

As part of the RS dissemination procedure, it should be taken into account the establishment of a list of related publications, data sets and other related works, that could be disseminated together with the RS, or that could be deposited elsewhere. In the latter case, let us remark that the links among all these objects will facilitate research reproducibility 37,61–63.

Strong dissemination methods can include the deposit of the RS in places like the *Agence pour la protection des programmes*[83] and the realization of Software Management Plans 40,66.

Finally, let us mention that it is the task of the policy makers to set the Open Science policies that should be applied in the evaluation procedure, as well as to establish the required dissemination level and related characteristics that are best adapted to the evaluation context at stake, as mentioned in the above considerations. Furthermore, evaluators should verify and check that the presented RS complies with the established requirements regarding free/open access and Open Science issues, as well as with the dissemination practices set by the policy makers.

---

[77]See for example the Directive 2009/24/EC of the European Parliament and of the Council from 23 April 2009 on the legal protection of computer programs, https://eur-lex.europa.eu/legal-content/EN/ALL/?uri=CELEX%3A32009L0024

[78]Please note that no license means All rights reserved.

[79] https://spdx.org/

[80]Free software is defined by the Free Software Foundation (FSF) in https://www.gnu.org/philosophy/free-sw.en.html

[81]Open source software is defined by the Open Source Initiative (OSI) in https://opensource.org/docs/osd.

[82] https://www.budapestopenaccessinitiative.org

[83] https://www.app.asso.fr/en

**(U) Use.** This is the step devoted to the more technical software evaluation issues. The goal of this **U**se step is not to propose some software performance evaluation estimations, as for RS the most important characteristic is its scientific correctness, but different levels of software quality can be taken into account here.

Evaluators considering a particular RS can have several approaches in mind. For instance, to check published results obtained through the use of the RS, in order to be able to find bugs or defects in the program that could affect the published results; to explore the scientific issues behind the RS computations (for example, to get better understanding of the implemented algorithms and the corresponding theoretical framework); to compare with other RS products; to take into account reproducible research issues 37,61–63 to measure the potential of the RS for the production of new scientific results...

In all these cases it is evident that the **U**se evaluation first goal should be to assess how much the RS team has facilitated its use, in concordance to the level of requirements set up by the policy makers. This basic step involves (*i*) checking how easy/difficult is to retrieve and install the RS, (*ii*) verifying if the RS has the necessary documentation and instructions to install, run and test the software, (*iii*) mentioning the requirements of the RS to some computing environments and to components that should be pre-installed, and (*iv*) providing the necessary examples and links to articles and other data, in order to help users to launch the first computations and to be able to verify and reproduce the already published results. In this regard, RS development teams preparing for a tough software examination procedure can find help in bibliographic references like 16,17,20,25,26,69–75 and the citations therein.

Evaluators should launch the RS, run and verify some examples and compare the output with the expected results. However, they must also have a look to the code. In our vision, running some examples do not dispense the evaluators from looking the RS source code. This calls for evaluation committees provided with some reasonable software skills. This **U**se evaluation basic step could be facilitated if the RS is installed in platforms like IPOL 37, where the software can be launched and the source code is also available for its study in the same platform. This **U**se step could also be helped if the software is declared validated by some well identified research community as seen in the above section Publication of research software.

But the notion of usability is also related to the concept of scientific validation (12, p.133) and also to software verification and validation, including the verification of the code correctness, clarity and simplicity 76. RS verification and validation and RS code correctness can be assessed, for example, by using software inspection techniques see 77 and also 78, with a comprehensive survey of "software inspection" (sometimes referred to as software review or software peer review) literature during 1980–2008. Correctness is the highest priority for scientific software, as scientists do science rather than software 21,76.

Thus, before applying this **U**se step, it is required to establish the level of exigence for the RS testing, that can go from launching a few examples, to verifying if the RS provides a test suite and has installed testing procedures 1, up to inspecting carefully each line of the code.

As mentioned above, other points to be evaluated concern the level of documentation of the RS, the management of versions, the portability, bug tracking, user interfaces, governance, user support... 25. In this **U**se step, software assessment procedures can go quite way up, including, for instance, the verification of the application of ISO/IEC standards[84]. See also 57 for a discussion on FAIR software and the Software Seal of Approval.

Finally, another issue that could be taken into account here concerns the relevance of the involved RS for technology transfer and industrial applications.

Last, let us recall that policy makers should set the definition of good/validated/successful RS that should be applied in the corresponding evaluation context, and they should as well indicate the expected level of reasonable/good/best software development practices. Besides, it is recommended that RS evaluators consider setting an evaluation matrix taking into account the different software aspects to be evaluated and the rate scale to be applied in each case.

**(R) Research.** This is the last step, the one where the standard research evaluation issues are to be taken into account 8,50,64,65, the place where the RS scientific value is to be assessed. Hence, it is also the point where the scientific software ecosystem requires full consideration 3 and where the related RS publication(s) that appear in the RS definition given in the definition dedicated section come into play, as their number, quality and impact reflects the quality and impact of the RS.

As this is the step for the evaluation of the research carried out with the RS and that is published in the related articles, the evaluation of both, software and articles, can be confounded in this global vision. Nevertheless, it is the intention of our proposition to put the RS at the center of the research evaluation that is under consideration, and policy makers and evaluators should decide the right balance between associated articles and software evaluation.

Moreover, this step includes the evaluation of the difficulty of the addressed scientific problems, the quality of the obtained results and theories, the efficiency of the proposed (and coded) algorithms, the participation in funded projects, the measure of the potential of the RS for the production of new scientific results, etc.

We consider that the basic level of this **R**esearch step should rely on the number and quality of the related publications, and on the number of their citations. In fact, the dissemination levels of such documents provide indications about up to what point the RS is being widely adopted by the scientific community, yielding, therefore, an estimation of the impact of the whole research work (articles and software).

---

[84]See for example ISO/IEC 9126 Software engineering - Product quality at https://en.wikipedia.org/wiki/ISO/IEC_9126.

Nowadays, citation of software alone is still not a fairly well adopted behavior by the scientific community, so we must rely on the citations of the publications directly related to the RS. As a rough approximation to quality, we can measure publications' impact through the number of citations. Indeed, as concluded in 79, *"citation numbers approximate with good accuracy the perceived impact of scientific publications"*. On the other hand, as we can see in 80, *"results in both studies [...] indicate that papers with code available online are more highly cited than those without"*. Thus, impact of some RS and impact of its related articles are, again, confounded.

In this **R**esearch step we can also take into account another evaluation item, namely to consider the estimated number of RS users. A widely used software product can also attract new funding and new collaborations 12 (point about quality and evaluation):

> *"La qualité d'un logiciel peut se mesurer par celle des articles associés au logiciel, mais aussi par le nombre d'utilisateurs qu'il est capable d'attirer, de collaborations et de contrats qu'il est capable de générer."*
>
> *[The quality of a software can be measured by the quality of its associated articles, but also by the number of users that it is able to attract, and the collaborations and contracts that it is able to generate.]*

Note that the citation number of a publication provides somehow an estimation of the number of *its users*, that is, the readers of the publication. Likewise, the citations of the RS related publications can provide another (rough) measure of the number of RS users. Moreover, a RS can also be included as a component in other RS, but this kind of impact is difficult to assess as, again, we cannot rely on pure RS citation issues yet.

Thus, the basic level of this **R**esearch step should rely on the number and quality of the related publications, the estimation of the number of their citations and the estimation of the RS citations (as a research output or as an included component) whenever possible. Higher levels of research quality evaluation should assess up to which point the RS is widely adopted by the scientific community and the impact of the whole research work (articles and software).

Similarly, to the previous **U**se step, the policy makers should set the definition of good/validated/successful RS that would be applied in the corresponding evaluation context as well as to indicate the expected level of reasonable/good/best research practices. Besides, it is recommended that RS evaluators set an evaluation matrix taking into account the different research aspects to be evaluated and the rate scale to be applied.

### 4.3 CDUR: a use case step-by-step
In order to facilitate its adoption, we revisit here the steps involved in the CDUR protocol in the following hypothetical evaluation instance.

In our example, the evaluation context corresponds to the recruitment, for a five-year research project, of a researcher with some RS experience. Within the funded project context, publications and some RS developments are expected as outputs that will be disseminated following Open Science best practices. The project funders set an evaluation committee with experts in both, the research topic at stake and software skills, to examine the CVs and the production presented by the candidates. The project grant funders also set that the balance between candidates' research and software competency should bias to stress research knowledge and experience, but also that RS development experience is to be required for all admissible candidates. The *quality evaluation method* will be applied, although a few impact stats could be handled by the committee, if available for all the candidates at a comparable level. Candidates are expected to have participated to the development of a RS at least as minor code contributor, and as a co-author of at least one associated research paper. The committee will set basic/good/excellent criteria for the CDUR procedure, in agreement with the project funders.

Following the grant project funders' instructions, the evaluation committee will analyse in detail just one RS and its associated articles for each candidate, and candidates are expected to signal in their CVs which is the work that is to be more thoroughly analysed by the committee. Then, the evaluation committee proposes to the grant project funders the following checklist for the application of the CDUR protocol to each RS:

**(C) Citation.**

1. Basic level: the RS has a well-established reference that identifies correctly the RS, its authors, the last version and related date. The role and the contribution of the candidate is well described.

2. Good level: the RS also gives a clear list of references to other RSs that have been included or that are necessary to run it.

3. Excellent level: the RS identification includes a complete metadataset or a software paper that complete the initial RS reference.

**(D) Dissemination.**

1. Basic level: the RS is available for the committee.

2. Good level: the RS is disseminated under a FOSS license following a procedure like the ones proposed in 14,24.

3. Excellent level: the RS has a SMP available for the evaluation committee 40,66.

**(U) Use.**

1. Basic level: committee members are able to launch the RS and test the provided examples without difficulty. The published results can be reproduced.

2. Good level: the code identifies well the scientific work and the implemented algorithms, giving the necessary references.

3. Excellent level: the code has good management practices that include a good documentation, testing procedures and version management.

**(R) Research.**

1. Basic level: the RS and the associated articles signed by the candidate are of good quality. The candidate is a minor code contributor and has reasonable software development experience.

2. Good level: the RS has been used to produce several articles and has been the object of a funded project. The candidate is a major code contributor.

3. Excellent level: the RS is used by other research teams and has been cited by their publications. The candidate is a major code contributor and/or the responsible of the RS project.

Finally, after having validated this checklist with the project grant funders, the evaluation committee starts to analyse the different candidates' work to obtain relevant information about their RS experience.

### 4.4 The CDUR protocol: final reflections
As we have already seen in the extended presentation of the CDUR procedure, each step in the protocol proposes to consider different elements of achievement. Each of these elements can reach different levels and the corresponding scale is to be set up by the policy makers considering a particular evaluation event. Thus, our protocol can be easily adapted to different circumstances: career evolution, recruitment, funding, RS peer review or other procedures to be applied by universities and other research institutions, research funders, or scientific journals, and it can also be adapted to different situations arising in different scientific areas. As mentioned before in the detailed explanation, each CDUR step is associated to some RS important issues:

**(C) Citation.** This step considers the citation issues that require setting a reference, and the identification of RS authors. The legal issues that appear in here correspond to the intellectual property associated to the authors and their affiliations.

**(D) Dissemination.** If the RS is disseminated, it should be under a correct free/open source software license, and following best dissemination practices. This is the step where Open Science issues are most relevant.

**(U) Use.** This step is devoted to software use and correctness, and it is also the step associated to reproducibility issues. It can be enhanced with best software practices.

**(R) Research.** This is the step associated to research quality and its impact.

In CDUR, each of these issues has been put in a particular place in the whole protocol, that is to be applied as a set of chronologically ordered steps.

Let us remark here that the CDUR protocol is clear in discriminating the diverse roles of the evaluators, the policy makers and the evaluated researchers, and identifies the level of policies and the different actions to be put into place by the different key evaluation actors.

In CDUR, policy makers should determine the relevance and the balance between the pure technical software aspects and the research aspects. RS development can involve high levels of both, technical software skills and research expertise, and in our view, it is important to recognise when the evaluation process is dealing with software skills or with research. For example, it is not the same to detect poor software practices in the recruitment of a mathematics researcher than to detect that the produced RS presents a severe lack of correctness. Besides, if good citation practices are missing, this matter could be easily improved, mostly if candidates know in advance which will be the good practices that are to be taken into account.

So, in CDUR not only each step can be applied with flexibility, but also the whole protocol can be adapted to different situations by the evaluation committees, following the policy makers' requirements. The only drawbacks that we have found are the necessary transparency in the establishment of the applied protocol as well as the necessity of balanced committees with both research and software skills. On the other hand, to raise these drawbacks and to understand when they are relevant in the evaluation protocol also helps to tackle them correctly, which becomes another of the CDUR's benefits.

Finally, to complete the list of the protocol advantages, let us consider CDUR in the context of the *world brain vision* of 11 (chapter 2). This chapter lists ten principles that can guide the future of scholarly communication. But we note that, when comparing article and RS dissemination, a fundamental difference arises. Up to now, RS dissemination is predominantly in the hands of the RS producers, a rather different key actor than those ruling journals in scholarly publishing. This is one of the reasons why sound evaluation procedures are capital for the progress of RS issues in the scientific ecosystem. Consequently, it is our belief that the adoption of evaluation protocols like CDUR will contribute to support the above mentioned principles, such as:

- maximize RS accessibility and usability,

- support and expand range of contributions with equity, diversity and inclusivity criteria,

- support community building, and

- promote high-quality research with heightened integrity;

which will have repercussions in the whole scholarly communication system.

## 5 Conclusion
In this paper we have analysed the concepts of research software, its authors, and the issues related to RS publications, referencing and citing. Then we have studied the evaluation issues such as the existing methods and its key actors. Regarding more specifically RS evaluation, we have detailed the ideas around the concept of successful software and its value scale in a scientific community. These preliminary steps open the path to the proposition of the CDUR protocol for RS evaluation, that comprises four steps dealing with **C**itation, **D**issemination, **U**se and **R**esearch, and that are to be applied in this chronological order. This protocol and its advantages have been thoroughly investigated, including the different actions and decisions of the key actors (policy makers, evaluators, evaluated researchers) and the wide flexibility for its application in several contexts.

Research software production is already part of the daily activities of many researchers, but it is still not sufficiently recognized in the research evaluation procedures that are being currently applied in the scientific world, as far as we know. On the other hand, the difficulties of scientists and engineers in finding software of their interest have been studied in 4,5 and involve a collection of serious drawbacks affecting RS development such as, for example, work duplication, reduced scientific reproducibility and poor return of funding agencies' investment. Indeed, RS limited visibility means that incentives to produce high-quality, widely shared, and codeveloped software may be lacking.

Thus, we consider that it is in the interest of the research communities and institutions to adopt clear and transparent procedures for the evaluation of research software. Procedures like the proposed CDUR protocol facilitate RS evaluation and will, as a consequence, improve RS sharing and dissemination, RS citation practices and, thus, RS impact assessment. This is an important step for the recognition of RS production and, therefore, to help scientists towards better, more efficient research.

*"Clearly, a policy is only as good as its enforcement"* 4 (p.15). Procedures such as CDUR are exigent regarding transparent decisions. When seeking quality results, it is generally advisable to avoid having social factors to take a relevant role in the evaluation process. Indeed, social evaluation methods could be as good as any other, if they finally happen to lead to a similar level of quality in the evaluation results, as with the qualitative or quantitative methods. But the social methods should be applied quite consciously rather than unconsciously, and the moment when these practices come into play should be clearly detected and highlighted. The main drawback is that, in this case, it is usually difficult to refer to transparent policies and decisions. In other words, when transparency is at stake, social influence in evaluation procedures should be neutralized, as seen in the evaluation section.

Many of the RS points discussed here have common issues with research data evaluation. For example, as we remark in 14, research data and RS could be disseminated following the same procedure. Therefore, it is easy to conceive a similar CDUR evaluation protocol for research data, suitably modified to take into account some of its specific features, mainly by adapting the **U**se step to data use. The other steps, **C**itation, **D**issemination, and **R**esearch can have a pretty similar presentation by changing RS for research data, but taking into account the fundamental differences that appear between software and data legal issues.

As proclaimed in 11, the evaluation of research is the keystone for the evolution of the Open Science policies and practices. It is our belief that research evaluation is also the keystone for the evolution of research software practices and for the full consideration of its role towards a more efficient science.

As a final conclusion, we hope that the adoption of protocols such as CDUR will motivate and consolidate evaluation policies and practices. This article wants to be a call for action to foster a debate on RS evaluation protocols. A debate which, possibly, will also require a careful observation of the practical applications of such protocols and to analyse their adjustment to a scientific landscape in constant and fast evolution. Yet, as warned in 10 (p.2), *"Enacting any new policy that requires a shift in culture also will require community support for successful and efficient implementation"*. Evolutions will come, some are already there. Therefore, the future roles of the RS key evaluation actors are likely to evolve and change current practices, which carries out challenges and opportunities. The RS roles could evolve as part of the whole scholarly communication or on their own, probably both at the same time. There will be movement backwards and forwards, but, in our view (and in agreement with the EC Expert Group report 11) the success of the foreseen evolutions will be associated to initiatives that put researchers' aims at the center of the various interests. Success will come if the different actors participating to build the future will work closely with the researchers, to create and to provide procedures and services that are valued and trusted by them.

## Data availability
### Underlying data
All data underlying the results are available as part of the article and no additional source data are required.

### Acknowledgements
This work is partially in debt to the research software producers of the Gaspard-Monge computer science laboratory (LIGM) at the University of Paris-Est Marne-la-Vallée, where the RS production has been studied by the first author since 2006[85]. Authors also acknowledge referees' suggestions and comments as they have helped to clarify and improve several parts of this work.

---

[85]http://igm.univ-mlv.fr/~teresa/2013octPostersAERES/PatrimoineLogicielLIGM/LogicielsLIGMPlume2013_EN.pdf

# F1000Research

# Open Peer Review

## Current Peer Review Status:

---

**Version 2**

Reviewer Report 27 November 2019

https://doi.org/10.5256/f1000research.23236.r57114

**Francisco Queiroz**
School of Design, University of Leeds, Leeds, UK

The authors have provided a revised article addressing the most important issues identified by the previous report. Overall, the improvement of Figure 1, the inclusion of the use case, and the call to action have made the article much more comprehensible and engaging – in which case I'd recommend its approval without any reservations.

***Competing Interests:*** No competing interests were disclosed.

***Reviewer Expertise:*** Digital and interactive design; UX; UI; Scientific Software usability

**I confirm that I have read this submission and believe that I have an appropriate level of expertise to confirm that it is of an acceptable scientific standard.**

---

**Version 1**

Reviewer Report 23 September 2019

https://doi.org/10.5256/f1000research.21946.r52525

**Jean-Pierre Merlet**
HEPHAISTOS Project, Inria, Inria, Sophia-Antipolis, Valbonne, France

This paper gives a good definition of what is a research software (RS) and I agree with the authors that an extensive version of what is a RS. I have however various remarks:

- A career of a researcher has extensively changed in the recent years: more chairs available for a relatively short period of time (typically 5 years) are given so that in this limited amount of time a relatively young researcher cannot afford to stand by the dissemination and maintenance work required by the evaluation scheme provided by the authors

- The authors still favor a specific type of RS, with availability, documented, licensed and with versioning. This is dangerous as many RS do not enter in this category: software that is produced to check a specific scientific result, software specifically designed to manage a particular system etc.. These software are not intended to be disseminated either as they are used only for checking and benchmarking or because they are very specific (for example a real-time software where the purpose is to save as much as possible computation time so that it cannot be general by essence as any general procedure will require tests that will penalize the performances) and therefore cannot be reproducible. As developing software is time intensive having too strict rule for the evaluation of software (such as documentation and availability) may discourage researchers producing them  although they provide some good idea of the efficiency of the underlying theoretical work.

- Citations of publications as a measure of impact is already questionable and I will say this is even worse for RS. It is quite current in the computer science community to say that the only good software is the one produced by the authors so that citations is limited and usually not very positive.

- I fully agree that software development may be a major component in the daily activity of a researcher and therefore must be taken into account in our evaluation. However the specificities of the RS and of the domain must also be taken into account in this evaluation. The CDUR procedure proposed by the authors is very fine but can be applied only on a limited number of RS types and his application will be disastrous for other types. For example it is easy to find RS for which the CDU part is not relevant while the R part is the only one of importance: for example a RS that performs the control of a specific robotic system cannot be reproducible (unless the reviewer has at hand an exactly similar robot), its use is limited to a specific prototype (and the RS has been designed for it) and has to bey tailored to manage another system so it cannot be disseminated. Hence the only evaluation criteria is the R part which is illustrated by the performance of the whole system, being given the performance of the hardware.

- I recognize that the authors are right in term of objectives: our current approach of RS development may lead to a waste of time with researchers programming again and again the same algorithms instead on focusing on their specific objectives. Some kind of mutualization will be beneficial provided that he RS is sufficiently disseminated and open to the community. But a researcher has to find the right balance between his/her research activity and the time devoted to maintenance, documentation and dissemination, a balance that greatly depends upon the domain. Furthermore good software practices for development is also dependent of the domains.

In summary although the authors have been very careful in their definition and evaluation rules and the absolute necessity of flexibility in the evaluation rules, they ended up with an evaluation procedure that may lead researcher to avoid going into the business of RS development. But this is nice paper that is worth being indexed for opening a discussion on RS evaluation according to the specific domain and context in which it has been developed.

**Is the work clearly and accurately presented and does it cite the current literature?**
Yes

**Is the study design appropriate and is the work technically sound?**
Yes

**Are sufficient details of methods and analysis provided to allow replication by others?**
Yes

**If applicable, is the statistical analysis and its interpretation appropriate?**
Not applicable

**Are all the source data underlying the results available to ensure full reproducibility?**
No source data required

**Are the conclusions drawn adequately supported by the results?**
Partly

***Competing Interests:*** No competing interests were disclosed.

***Reviewer Expertise:*** computer science, robotics,mathematics

**I confirm that I have read this submission and believe that I have an appropriate level of expertise to confirm that it is of an acceptable scientific standard.**

Reviewer Report 13 September 2019

**Francisco Queiroz**
School of Design, University of Leeds, Leeds, UK

Overall, this is a clear, well-written article outlining a novel procedure that could impact, in positive ways, policies and practices regarding scientific software development and impact. However, I believe there are some aspects of the article that could be either improved or clarified.
   1. The study seems to be mostly based on secondary research – more specifically, arguments and ideas are developed based on findings from literature. Moreover, it can be assumed, the study is informed by the proximity of the authors to the academic/professional environment it investigates. However, there is no clear indication on how the material that serves as basis for the analysis was selected (indexed databases? specialized resources such as publications and conferences?), in which case it would be interesting to include a justification on why those sources are relevant.

2. Additional material that could, potentially, be referenced are Chue Hong`s position paper on the need for a framework for comparing software[1]; and also ESRC's plans of developing a software accreditation framework[2].

3. In section 2, the authors write: "Note that to cite software in science is different to citing scientific (or more precisely, research) software, which is the issue here". Perhaps the difference could be clarified?

4. In 2.1, the authors write: "in 12,13 we can find the following definition". Given that the definition`s authorship is originally from 12, it would be probably more appropriate to write something along the lines of "a definition elaborated by 12 was summarized by 13 as"

5. There are several passages in both French and English. Wouldn't it be enough to keep the English versions (along with an indication that they were translated from French by the authors?)

6. The authors write: "Finally, let us mention that, as in the case of publications, the research software production of a laboratory is decided and proposed by the lab's members, and it is approved by the leading institutions during the usual laboratory evaluation and funding procedures". Is that true for every case of research software production and are there any references supporting that?

7. Figure 1, illustrating concepts appearing in the study, could be improved, possibly by matching label colors to the colors of the rectangles for easier identification.

8. Is not clear why there are two screenshots of Plume (one in French and one in English). One would probably be enough?

9. The authors write: "(...) we remark that there can be several classic articles associated to a single RS, and that could be used as its reference". Could references to some of those classic articles be provided?

10. The study ends without (i) considerations on future research, (ii) clear indications on initiatives/pilots for the procedure, (iii) a call to action for the community to test it. Overall, it's not clear what the next steps are.

11. Regarding the applicability of the procedure, the article would benefit from including a table or diagram (probably in Section 4) explaining the procedure, preferably featuring a clear, step-by-step description of the process. That should clarify how CDUR could be adopted, increasing the potential for its adoption and use.

Experiences (WSSSPE4). *University of Manchester, CEUR Workshop Proceedings*. 2016.
2. EPSRC Software Infrastructure strategy 2018. *EPSRC*. 2018. Reference Source

**Is the work clearly and accurately presented and does it cite the current literature?**
Yes

**Is the study design appropriate and is the work technically sound?**
Partly

**Are sufficient details of methods and analysis provided to allow replication by others?**
Partly

**If applicable, is the statistical analysis and its interpretation appropriate?**
Not applicable

**Are all the source data underlying the results available to ensure full reproducibility?**
No source data required

**Are the conclusions drawn adequately supported by the results?**
Yes

***Competing Interests:*** No competing interests were disclosed.

***Reviewer Expertise:*** Digital and interactive design; UX; UI; Scientific Software usability.

**I confirm that I have read this submission and believe that I have an appropriate level of expertise to confirm that it is of an acceptable scientific standard, however I have significant reservations, as outlined above.**

# Comments on this article

Version 2

Author Response 19 Oct 2020
**Teresa Gomez-Diaz**, University of Paris-Est Marne-la-Vallee, Marne-la-Vallée, France

While doing a presentation of this work at the University of Cantabria (Spain) in September 2020, the former Professor of this University Carlos Ruiz de Velasco reminded us of the importance of using the right data structures in research software (and in software in general).

This is why the final version of the slides (in Spanish), available at

http://igm.univ-mlv.fr/~teresa/logicielsLIGM/conferenciasSeptiembre2020/20200903_CDUR_Final_ES.pdf

include data structures at the (R) research step of the CDUR procedure (slide 14/15).

This also implies some corrections in the CDUR procedure proposed in this paper, as algorithms and data structures should be considered at the same level and at the same time, mainly in the (U) Use and the (R) Research steps.

With many thanks to our Professor, colleague and friend Carlos Ruiz de Velasco.

***Competing Interests:*** No competing interests have been detected.

---

The benefits of publishing with F1000Research:

- Your article is published within days, with no editorial bias

- You can publish traditional articles, null/negative results, case reports, data notes and more

- The peer review process is transparent and collaborative

- Your article is indexed in PubMed after passing peer review

- Dedicated customer support at every stage

For pre-submission enquiries, contact research@f1000.com

F1OOO Research